 \documentstyle[12pt]{article}
 
\begin{document}

 \def\to{\rightarrow}

 \rightline{EFI 93-44}

 \bigskip \medskip \begin{center} \large {\bf The Spectral Problem for
 the \\
 ${\bf q}$-Knizhnik-Zamolodchikov Equation \\
 and Continuous q-Jacobi Polynomials} \\

 \bigskip \bigskip \normalsize

 Peter G. O. Freund\footnote{Work supported in part by the NSF:
 PHY-91-23780} and Anton V. Zabrodin\footnote{Permanent address:
 Institute of Chemical Physics, Kosygina Str. 4, SU-117334, Moscow,
 Russia }\\ {\it Enrico Fermi Institute, Department of Physics \\ and
 Mathematical Disciplines Center \\ University of Chicago, Chicago, IL
 60637} \\

 \end{center}
 \bigskip
 \centerline{ABSTRACT}
 \begin{quote}

The spectral problem for the q-Knizhnik-Zamolodchikov equations for
$U_{q}(\widehat{sl_2}) (0<q<1)$ at arbitrary level $k$ is considered.
The case of two-point functions in the fundamental representation is
studied in detail.The scattering states are given explicitly in terms
of continuous q-Jacobi polynomials, and the $S$-matrix is derived from
their asymptotic behavior. The level zero $S$-matrix is shown to coincide,
up to a trivial factor, with the kink-antikink $S$-matrix for the
spin-$\frac{1}{2}$ XXZ antiferromagnet. In the limit of infinite level we
observe connections with harmonic analysis on $p$-adic groups with the prime
$p$ given by $p=q^{-2}$.

 \end{quote}
 \newpage

 \bigskip
 \bigskip
 \leftline{\bf 1.  Introduction}

 There is accumulating evidence for the idea \cite{FZK} that excitation
 scattering in integrable models is ``geometric", i.e., that the
 corresponding wave functions are spherical functions of certain quantum
 symmetric spaces.
 With this idea in mind, we have recently derived \cite{FZ} the physical
 $S$\--matrix for the scattering of kinks and antikinks
  for the (spin\--$\frac{1}{2}$)
XXX and XXZ antiferromagnets, starting from the zero
 level  $q$\--Knizhnik\--Zamolodchikov (q\--KZ) equations
for $U_q(\widehat{sl_2})$
 in the fundamental representation (the Heisenberg XXX case corresponds to
$q=1$). These $q$\--KZ equations \cite{FR}
 are q-deformations
 of the ordinary first order differential KZ equations \cite{KZ}, which,
 in turn, are similar to the
 familiar Dirac and Bargmann-Wigner \cite{BW} equations. That the $q$-KZ
equations
apply in the kink-antikink problem has to do with the fact that kinks
and antikinks are known to be spin $\frac{1}{2}$ excitations \cite{FT, KR}.
 Here we extend this work to nonvanishing values of the level $k$.
 This brings the continuous $q$\--Jacobi polynomials into play and
 physics\--wise concerns $SL_q (k+2)$\--magnetics (or the corresponding
 generalizationis of Baxter's eight vertex model).
 The first\--order $q$\--KZ operator is cast here in the r\^{o}le of the
 radial part of a Dirac operator, whose ``square" yields the radial part
 of the Laplace operator on the quantum symmetric space.
 This picture leads to useful insights concerning the
 $p$\--adics\--quantum\--group connection \cite{M, F, Z, FZCO}.

 As in \cite{FZ}, we deal here with the {\em spectral} problem for the q\--KZ
 operator and not the monodromy problem considered by others \cite{FR}.
 The difference between these two problems will be discussed: in a certain
 sense they are each other's duals.

 The paper is organized as follows.
We start in Section 2 with the classical ($q=1$) case where the
q\--KZ equations are the usual differential
KZ equations \cite{KZ} written with trigonometric rather than rational
classical $r$\--matrix. In other words, one uses a
Borel type polarization instead of a parabolic polarization.
The KZ equation in trigonometric form contains the highest weight of the
vacuum representation which, when continued to the whole complex plane, may
be viewed as a spectral parameter $\lambda$.
The {\em spectral} problem yields a non\--trivial $\lambda$\--dependent
$S$\--matrix
which in case of level zero coincides (up to a constant matrix factor) with
the kink\--antikink $S$\--matrix in the spin $\frac{1}{2}$ isotropic (XXX)
Heisenberg model, $\lambda$ being the relative rapidity of the excitations.

In Section 3 we solve the spectral problem for the $q$-deformed KZ equations
at arbitrary level $k$.
The scattering states are explicitly found in terms of continuous
$q$\--Jacobi polynomials.
Section 4 deals with the special case $k=0$ where the obtained $S$\--matrix
coincides, up to a trivial matrix factor, with the kink\--antikink
$S$\--matrix in the spin$-\frac{1}{2}$ XXZ antiferromagnet.
The corresponding scattering states may be interpreted as spinorial
harmonics on the {\em quantum} $SL(2)$ group.
Another interesting special case is the limit of infinite level
($k \rightarrow \infty$) considered in Section 5.
If $q^2 = p^{-1}$ and $p$ is a prime number this case turns out to be
closely connected with harmonic analysis on the $p$\--adic group
$PGL(2, {\bf Q_p})$.
We discuss some aspects of this connection and suggest an ``arboreal''
interpretation of spinorial harmonics on the $p$\--adic group in terms of
Bruhat\--Tits trees.
Section 6 contains a general discussion and conclusions.
In Appendix A some technical details related to Section 3 are explained.
Appendix B contains a brief review of the continuous $q$\--Jacobi
polynomials.

 \bigskip

 {\bf 2. The classical (q=1) KZ equation, its spectral problem and $S$-matrix.}

 We start from the classical $(q = 1)$ KZ\--equation for
 $\widehat{sl_2}$ in the fundamental representation for level $k$.The case
 $k=0$ was treated in our earlier paper \cite{FZ}, here we right away address
 the case of generic $k$.
 With normal
 ordering relative to a Borel polarization, consider the matrix element
  $$ \Psi (x_i) = \langle \Omega ' | \Phi (x_2) \Phi (x_1) |
 \Omega \rangle \eqno(2.1) $$
 of the product of two vertex operators $\Phi$ between suitable vacuum
 states.This matrix element depends on only one variable, which in an
 {\em additive} parametrization can be chosen as
 $$ x = {1 \over 2}(x_1 - x_2) ~.  \eqno(2.2) $$
  The matrix element $\Psi (x_i)$ is then a
 ${\bf C}^2 \otimes {\bf C}^2$ valued function $\Psi(x)$ of the
 variable $x$.  This $\Psi (x)$ obeys the KZ equation, which, for
 $\widehat{sl_2}$ at level $k$ and in the fundamental representation,
 takes the form $$ (k+2) \frac{d \Psi (x)}{dx} = (r_{12} (x) + \pi_1 (H))
 \Psi (x) ~, \eqno(2.3) $$ where $r_{12} (x)$ is the familiar
 trigonometric solution of the {\em classical} Yang\--Baxter equation
 $$
r_{12} (x) = (\coth x) [ E \otimes F + F \otimes E
 + {1 \over 2}
  H \otimes H - {1 \over 2} {\bf 1} \otimes {\bf 1} ] - E \otimes F + F
 \otimes E
$$
$$
 \pi_1 (H) = i(k+2){\lambda H \otimes {\bf 1}, \\
  E = \left
 ( \begin{array}{c c} 0 &1 \\ 0 & 0 \end{array} \right ), F = \left (
 \begin{array}{c c} 0 & 0 \\ 1 & 0 \end{array} \right ),
  H = \left (
 \begin{array}{c c}
 1 & 0 \\
 0 & -1
 \end{array} \right ),
 \eqno(2.4)
$$
 with $\lambda$ the (real) parameter of our spectral
 problem.  In the combination $k+2$ on the left hand side of eq.~(2.3),
 the term 2 is the dual Coxeter number $g=2$ of $sl_2$.  The choice
 of the term proportional to the $4 \times 4$ unit matrix in the
 expression of $r_{12} (x)$ (the last term in the square bracket in the
 first equation (4)) is irrelevant as far as the Yang\--Baxter equation
 is concerned, but considerably simplifies the argument.  With respect
 to a basis $v_a \otimes v_b , a, b = \pm$ of ${\bf C}^2 \otimes {\bf
 C}^2$, we can expand $$ \Psi (x) = a (x) v_+ \otimes v_+ + f (x) v_+
 \otimes v_- + g (x) v_- \otimes v_+ +
  b (x) v_- \otimes v_- \eqno(2.5a) $$ In components the KZ equation (2.3)
 then becomes $$
  {{d} \over {dx}} a (x) = i \lambda a (x), ~~
  {{d} \over {dx}} b (x) = - i \lambda b (x) \eqno(2.5b) $$ $$ \left (
 \frac{d}{dx} + {1 \over {k+2}} \coth x - {1 \over {k+2}} \sigma_1 \coth x
 \right ) \psi (x) = i ( \lambda \sigma_3 - {1 \over {k+2}} \sigma_2 ) \psi
 (x), \eqno(2.5c)
 $$
 $$
  \psi (x) = \left ( \begin{array}{c} f (x) \\ g (x) \end{array}
 \right ).
\eqno(2.5d)
 $$
 The nonzero weight com\-po\-nents $a(x)$ and
 $b(x)$ de\-couple and are trivi\-al \\ (eqs.~(2.5b)).  Hence\-forth we
 ignore them.  The interesting equation (2.5c) involves the zero weight
 sector.  This equation is similar to a Dirac equation with our
 spectral parameter $\lambda$ playing the role of mass in a
 ``$\gamma_5$\--type'' mass term.  It is convenient to introduce the
 triplet and singlet combinations $$ F^+ (x) = f(x) + g (x), ~~ F^- (x)
 = f(x) - g (x) ~, \eqno(2.6) $$ which obey the first order differential
 equations $$ F^{- \prime } + {2 \over {k+2}} (\coth x ) F^- = (i \lambda
 - {1 \over {k+2}}) F^+, ~~ F^{+ \prime } = (i \lambda + {1 \over {k+2}} )
 F^- ~.  \eqno(2.7)
 $$ Here prime stands for derivative with respect to $x$.  The
 equations (2.7) lead to the decoupled system of second order
 differential equations
 $$
 \left ( {{d^2} \over {dx^2}} + {2 \over {k+2}}
 \coth x
 {{d}\over {dx} }
 \right ) F^+ (x) = - \left ( \lambda^2 + {1 \over {(k+2)^2}}
 \right ) F^+ (x)
 $$
 $$
 \left ( {{d^2} \over {dx^2}} + {2 \over{k+2}}
 \coth x {d \over
 {dx}}- {{2} \over {(k+2) \sinh^2 x}} \right ) F^- (x) =
 $$
 $$
   = - \left ( \lambda^2
 + {1 \over {(k+2)^2}} \right ) F^- (x) ~.  \eqno(2.8)
 $$

The first of these equations is a special case of a theorem of Matsuo
\cite{MA}. Both equations (2.8) can also be extracted from Cherednik's
papers \cite{CHE}, as has been pointed out to us by A. Veselov.
 It is convenient to introduce the new variable $z$, the new functions
 $G^\pm$, and the new parameters $n , \ell$ defined by

 $$
 \begin{array}{c c}
& z= \cosh x , \\
     & \ell = \frac{1}{k+2},
 \end{array}
 \begin{array}{l }
 G^\pm (z) = F^\pm (x) , \\
 n = - \ell + i \lambda
 \end{array}
 \eqno(2.9a)
 $$
 in terms of which eqs.~(2.8) become
 $$
 \left [
 \frac{d^2}{dz^2} +
 \frac{2 \ell +1}{z^2 -1} z \frac{d}{dz} \right ]
 G^+ (z) =
 \frac{n(n+2 \ell)}{z^2 -1} G^+(z)
 \eqno(2.9b)
 $$
 $$
 \left [
 \frac{d^2}{dz^2} +
 \frac{2 \ell +1}{z^2 -1} z \frac{d}{dz} \right ]
 G^- (z) = \left ( n (n + 2 \ell) + \frac{2 \ell}{z^2 - 1} \right )
 \frac{1}{z^2-1} G^- (z)
 \eqno(2.9c)
 $$
 Equation (2.9b) is the familiar differential equation obeyed by the
 Gegenbauer functions $C_n^\nu (z)$ for $\nu = \ell$ so that
 $$
 G^+ (z) = C_n^\ell (z)
 \eqno(2.10)
 $$
 with $n$ and $\ell$ given by eq.~(2.9a).
 In particular for $k=0$, so that $ \ell = {1 \over 2}$, the Gegenbauer
 functions reduce to Legendre functions in agreement with [2].
 Similarly $G^- (z)$ is also a Gegenbauer function, which in
 the same limit turns into an associated Legendre function (remember,
 these {\em too} are special cases of Gegenbauer functions).
 From the known asymptotics of Gegenbauer functions \cite{GEG}, we extract the
 $S$\--matrix.
 It is diagonal in the triplet/singlet $((+)/ (-))$  basis:
 $$
 S_{+}(\lambda) = \frac{c(\lambda)}{c(-\lambda)},  ~~
  S_{-}(\lambda) = \left ( \frac{\ell + i \lambda}{\ell -
i \lambda} \right )S_{+}(\lambda),
 \eqno(2.11a)
 $$
where the $c$\--function is
$$
c(\lambda) = \pi^{1/2}\frac{\Gamma(i \lambda)}{\Gamma(i \lambda + \ell)}.
\eqno(2.11b)
$$
Of special importance is the case $k=0$, which we treated in \cite{FZ}.
Here we shall indicate the $S$-matrix elements and $c$-function for this
$k=0$ case by using the superscript $(0)$.They are
$$
S_{+}^{(0)}(\lambda) = \frac{c^{(0)}(\lambda)}{c^{(0)}(-\lambda)},  ~~
S_{-}^{(0)}(\lambda) = \left (\frac{1 + 2i \lambda}{1 - 2i \lambda} \right )
S_{+}^{(0)}(\lambda),
\eqno(2.12a)
$$
with
$$
c^{(0)}(\lambda) = \pi^{1/2}\frac{\Gamma(i \lambda)}{\Gamma(i \lambda
+ \frac{1}{2})}.
\eqno(2.12b)
$$
As shown in \cite{FZ},
$S_{+}^{(0)}(\lambda)$ and $-S_{-}^{(0)}(\lambda)$ (the minus sign to be
discussed below) are the eigenvalues of the Heisenberg XXX model kink-antikink
scattering matrix.

We wish to draw attention here to two points.

First,as far as $sl_2$ representations are concerned, $\Psi(x)$, valued in
the tensor product of two two-dimensional representations, decomposes into
one singlet and three triplet components. Yet, as we saw, the two
non-vanishing weight triplet components decouple, leaving a two-dimensional
"spin-$\frac{1}{2}$"-like system obeying the Dirac-like equation (2.5).
A further study of this metamorphosis would be of interest.

Second, the steps leading from the coupled first-order system (2.5) to
the decoupled second order system (2.8), completely parallel those which
lead from the coupled first order Dirac equations to the eigenfunctions of
the decoupled second order Laplace operator. It will be worth keeping this
in mind, when we repeat these steps at the quantum level.

\newpage

 \leftline{\bf 3.  q-deformed case}

 Let us start by recalling some facts about q\--KZ equations for 2\--point
 functions in the $U_q ( \widehat{sl_2})$\--case $(0 < q < 1)$.
 We use the notations of \cite{DFJMN}, in a slightly modified form.
 Consider a correlation functions of two $q$\--vertex operators
 $$
 \Psi (z_1 , z_2) = \langle \Omega ' | \Phi (z_2) \Phi (z_1) | \Omega
 \rangle \in V \otimes V
 \eqno(3.1)
 $$
 where $V \cong {\bf C}^2$ is a linear space on which now 2\--dimensional
 representations of $U_q ( \widehat{sl_2})$ act.
 We fix a basis $\left \{ v_+ , v_- \right \}$ in $V$.
 With a proper definition of $q$\--vertex operators, $\Psi (z_1 , z_2)$
 depends only on $z_1 / z_2$ (we have now switched
 from the additive parametrization of Section 2, to a {\em multiplicative}
 parametrization), so we consider the function
 $$
 \Psi (z) = \langle \Omega ' | \Phi (z^{-1} ) \Phi (z) | \Omega \rangle ~~.
 \eqno(3.2)
 $$
 For level $k$ vertex operators, $\Psi (z)$ satisfies the first order q-KZ
 difference equation:
 $$
 \Psi (q^{k+2}z) = \rho (z) (q^{- \phi} \otimes 1 ) R (z) \Psi (z)
 \eqno(3.3)
 $$
 where the $R$-matrix $R(z)$ is defined by explicit action in $V \otimes
 V$ as follows:
 $$
 R(z) v_\pm \otimes v_\pm = v_\pm \otimes v_\pm
 \eqno(3.4a)
 $$
 $$
 R(z)v_+ \otimes v_- =
 \frac{q (1-z^2)}{1-q^2z^2} v_+ \otimes v_- +
 \frac{(1-q^2)z^2}{1-q^2z^2} v_- \otimes v_+
 $$
 $$
 R(z)v_- \otimes v_+ =
 \frac{1-q^2}{1-q^2z^2} v_+ \otimes v_- +
 \frac{q(1-z^2)}{1-q^2z^2} v_- \otimes v_+
 \eqno(3.4b)
 $$
 The operator $q^{- \phi}$ acts on the basis vectors by multiplication:
 $$
 q^{- \phi}v_\pm = q^{\mp 2i \lambda} v_\pm
 \eqno(3.5)
 $$
 where $\lambda$ is a spectral parameter.
 In the q\--KZ equations considered in \cite{DFJMN, FR} $\lambda$ takes a
 particular value depending on the choice of vacuum states $\Omega ,
 \Omega '$ in (3.2).
 When we are interested in the spectral problem for the difference
 operator in (3.3) (rather than monodromy properties of the solutions)
 $\lambda$ plays the role of spectral parameter (this becomes evident from
 eq.~(3.10) below).
 We have introduced $i$ in (3.5) (just like in eq.~(2.5c) in the classical
 case) so that the continuous spectrum will correspond to real values of
 $\lambda$.

 Finally, $\rho(z)$ in (3.3) is the scalar multiplier defined in
 \cite{DFJMN}:
 $$
 \rho (z) = q^{- 1/2} \cdot
 \frac{(q^2 z^2; q^4)_\infty^2}{(z^2; q^4)_\infty (q^4z^2; q^4)_\infty }
 \eqno(3.6)
 $$
 where the standard notation    \cite{GR}
 $$
 (z; q)_n = \prod_{j=0}^{n-1} (1 - zq^j) ; ~~~
 (z; q)_\infty = \lim_{n \rightarrow \infty} (z; q)_n
 \eqno(3.7)
 $$
 is used.
 It is shown in \cite{FR} that this multiplier comes from restriction of
 the universal $R$\--matrix for $U_q ( \widehat{sl_2})$ to the tensor
 product of two 2\--dimensional representations.
 Though crucial in the monodromy problem, $\rho (z)$ is irrelevant for our
 purposes here, because one can gauge it away without altering the
 spectral properties.

 Due to the specific form (3.4) of the $R$\--matrix, the $v_+ \otimes v_+$
 and $v_- \otimes v_-$ components of $\Psi (z)$ decouple, and each of them
 obeys a scalar first order difference equation as in the classical case
 (2.5b).
 Again, the non\--trivial equations come from the zero\--weight sector of
 the $R$\--matrix (3.4).
 After reduction to the zero\--weight subspace of $V \otimes V$ we obtain
 for the two components of
 $$
 \psi (z) \equiv f(z) v_+ \otimes v_- + g(z) v_- \otimes v_+ ,
 \eqno(3.8)
 $$
 the following system of difference equations
 $$
 f \left ( q^{k+2}z \right ) = q^{-2i \lambda}
 \frac{q(1-z^2)}{1-q^2z^2}
 f(z)+q^{-2i \lambda}
 \frac{1-q^2}{1-q^2z^2} g (z)
 $$
 $$
 g \left ( q^{k+2}z \right ) = q^{2i \lambda}
 \frac{(1-q^2)z^2}{1-q^2 z^2}
 f(z)+q^{2i \lambda}
 \frac{q(1-z^2)}{1-q^2z^2} g(z) ~~.
 \eqno(3.9)
 $$
 There is a more suggestive form of (3.9) which resembles the classical
 spectral problem (2.7).
 Calling $R_0(z)$ the zero\--weight part (3.4b) of the $R$\--matrix and
 introducing the diagonal $2 \times 2$ matrix $\Lambda = diag \left \{
 q^{2 i \lambda} , q^{-2i \lambda} \right \}$, one can rewrite (3.9) in
 the form
 $$
 T^{-1} R_0 (z) \psi (z) = \Lambda \psi (z)
 \eqno(3.10)
 $$
 where $T$ is the shift operator:  $T \psi (z) = \psi (q^{k+2}z)$.
 This equation does look like a finite\--difference analog of (2.7).

 Guided by the classical limit we interpret (3.10) as the radial part of a
 discrete ``Dirac\--like" equation for a particle on a curved quantum space.
 Let us rewrite (3.9) in terms of the discrete ``radial coordinate" $n$
 which we assume to be a non\--negative integer.
 To do this, it is convenient to redefine the parameters as
 $$
 p = q^{k+2}
 \eqno(3.11)
 $$
 $$
 \ell = \frac{1}{k +2}
 \eqno(3.12)
 $$
 $$
 u = 2 \ell \lambda
 \eqno(3.13)
 $$
 so that $q = p^\ell$.
 Setting $z = p^{n+\ell}$ and calling $f(p^{n+ \ell}) = f_n$, $g(p^{n+l})
 = g_n$ we obtain the following system of recurrence relations:
 $$
 \left (
 \begin{array}{c}
 f_{n+1} \\
 g_{n+1}
 \end{array}
 \right )
 = (1-p^{2n+4 \ell} )^{-1}
 \left (
 \begin{array}{c c}
 p^{-iu+ \ell} (1-p^{2n+2\ell}) & p^{-iu} (1-p^{2 \ell}) \\
 p^{iu} (1-p^{2 \ell} ) p^{2n+2 \ell} & p^{i u + \ell} (1-p^{2n+2 \ell})
 \end{array}
 \right )
 \left (
 \begin{array}{c}
 f_n \\
 g_n
 \end{array}
 \right ).
 \eqno(3.14)
 $$

 By a straightforward but somewhat lengthy calculation it can be shown
 that the linear combinations
 $$
 F_n^\pm = f_n \pm p^{-iu}g_n
 \eqno(3.15)
 $$
 obey second order recurrence relations of the form
 $$
 \frac{1-p^{2n+4 \ell}}{1-p^{2n+2 \ell}}
 F_{n+1}^\pm +p^{2 \ell}
 \frac{1-p^{2n-2}}{1-p^{2n+2 \ell-2}} F_{n-1}^\pm =
 $$
 $$
 = \mp \frac{(1-p^2)(1-p^{2 \ell})p^{2n+2 \ell -2}}{(1-p^{2n+2
 \ell})(1-p^{2n+2l-2})}
 F_n^\pm + p^\ell (p^{iu} + p^{-iu})F_n^\pm
 \eqno(3.16)
 $$
 (Equations (3.15) and (3.16) are the quantum counterparts of the
 ``classical" equations (2.6) and (2.8)).
 In the Appendix A we show how one can come to (3.16) from (3.14) without
 knowing the combinations (3.15) a priori.

 A natural boundary condition is again the finiteness of $F_n^\pm$ at $n=0$.
 Note that due to the specific form of the coefficients in (3.16), we
 don't need to fix values of $F_n^\pm$ at {\em two} points (say, $n=0$ and
 $n=1$), which would be the usual thing to do for second\--order
 recurrence relations.
 In the case at hand the regularity at $n=0$ already determines the
 solutions up to arbitrary constant factors which, in their turn, are
 determined by the values of $F_1^\pm$.

 With these very boundary conditions,  the eqs.~(3.16) are the recurrence
 relations for certain
 $q$\--Jacobi polynomials.
 (see, for example, \cite{GR} and Appendix B).
 This observation allows us to solve (3.16).
 Specifically,
 $$
 F_n^+ = F_1^+ R_{n-1}^{(\ell - 1/2 , \ell + 1/2 )} ( \frac{1}{2} (p^{iu} +
 p^{-iu}); p)
 \eqno(3.17a)
 $$
 $$
 F_n^- = F_1^- p^{-n+1}
 \frac{1-p^{2n+2 \ell -1}}{1-p^{2 \ell +1}} R_{n-1}^{(\ell + 1/2 , \ell -
 1/2)}
 (\frac{1}{2} (p^{iu} + p^{-iu}); p)
 \eqno(3.17b)
 $$
 where $R_n^{(\alpha , \beta )} (x; p)$ are the {\em continuous q\--Jacobi
 polynomials}.
 For their definition and brief review of their properties see Appendix B.

 For real values of $u$ in the Brillouin zone $- \frac{\pi}{\log p} < u
 \leq \frac{\pi}{\log p}$ the wave functions (3.17) are the scattering
 states for the q-KZ operator (l.h.s. of eq.~(3.10).
 To see this, it is necessary to find the asymptotics of $F_n^\pm$, or in
 other words, of the continuous $q$\--Jacobi polynomials at large $n$.
 Fortunately, this is known \cite{GR}, so that we have
 $$
 F_n^+ |_{n \rightarrow \infty} = N_+F_1^+
 p^{n \ell - \ell}
 (p^{inu}c_+ (-u) + p^{-inu}c_+ (u))
 \eqno(3.18a)
 $$
 $$
 F_n^- |_{n \rightarrow \infty} = N_-F_1^-
 \frac{p^{n \ell - \ell +1}}{1-p^{2 \ell +1}}
 (p^{inu}c_- (-u) + p^{-inu}c_- (u))
 \eqno(3.18b)
 $$
 where (up to a non-essential constant)
 $$
 c_\pm (u) =
 \frac{p^{iu}}{1 \pm p^{iu+ \ell}} \cdot
 \frac{\Gamma_{p^2} (iu)}{\Gamma_{p^2} (iu + \ell )}
 \eqno(3.19)
 $$
 and the $q$-gamma function is defined as \cite{GR}
 $$
 \Gamma_q(x) =
 \frac{(q;q)_\infty}{(q^x; q)_\infty} (1-q)^{1-x} .
 \eqno(3.20)
 $$
 For the expressions of the constants $N_\pm$ and other details see
 Appendix B $(N_\pm = N_{\ell \mp 1/2 , \ell \pm 1/2}$ in (B11)).
 The expression (3.19) does indeed look like a superposition of incoming
 and outgoing waves.

 Now we are in a position to derive the $2 \times 2$ $S$\--matrix.
 Let us rewrite the asymptotics (3.18) in terms of the original components
 $f_n$ and $g_n$,
 $f_n = (1/2) (F_n^+ + F_n^-)$,
 $g_n = (1/2) p^{iu} (F_n^+ - F_n^-)$ (as follows from eq.~(3.15)):
 $$
 f_n|_{n \rightarrow \infty} =
 \frac{p^{n \ell - \ell}}{2 (1-p^{2 \ell +1})}
 \left \{
 \left (
 \frac{N_+F_1^+ (1-p^{2 \ell +1})p^{-iu}}{1 +p^{-iu+ \ell}} +
 \frac{N_-F_1^- p^{-iu}}{1-p^{-iu+ \ell}}
 \right ) c( -u)p^{inu}+
 \right .
 $$
 $$
 \left .
 + \left (
 \frac{N_+F_1^+ (1-p^{2 \ell +1})p^{iu}}{1+p^{iu + \ell}}
 + \frac{N_-F_1^-p^{iu}}{1-p^{iu + \ell}}
 \right ) c(u)p^{-inu} \right \}
 \eqno(3.21a)
 $$
 \bigskip
 $$
 g_n|_{n \rightarrow \infty} =
 \frac{p^{n \ell - \ell}}{2 (1-p^{2 \ell +1})}
 \left \{
 \left (
 \frac{N_+F_1^+ (1-p^{2 \ell +1})}{1 +p^{-iu+ \ell}} -
 \frac{N_-F_1^- }{1-p^{-iu+ \ell}}
 \right ) c( -u)p^{inu} +
 \right .
 $$
 $$
 \left .
 + \left (
 \frac{N_+F_1^+ (1-p^{2 \ell +1})p^{2iu}}{1+p^{iu + \ell}}
 - \frac{N_-F_1^-p^{2iu}}{1-p^{iu + \ell}}
 \right ) c(u)p^{-inu} \right \}
 \eqno(3.21b)
 $$
 where
 $$
 c(u) =
 \frac{\Gamma_{p^2}(iu)}{\Gamma_{p^2}(iu + \ell)} ~~~.
 \eqno(3.22)
 $$
 The $S$-matrix can be readily determined from (3.21) by imposing special
 boundary conditions at infinity rather than at the origin.
 Namely, suppose the incoming wave at infinity has the form
 $ \left (
 \begin{array}{c}
 1 \\ 0
 \end{array}
 \right) p^{inu+n \ell}$, i.e. the g-component is zero.
 Then the corresponding outgoing wave is generally of the form
 $\left (
 \begin{array}{c}
 S_{11} \\
 S_{21}
 \end{array}
 \right ) p^{-inu+ n \ell}$ where $S_{11}$ and $S_{21}$ are, by definition,
 the matrix elements of the $S$\--matrix.
 Similarly, for the incoming wave of the type $ \left (
 \begin{array}{c}
 0 \\ 1
 \end{array} \right ) p^{inu+n \ell}$ we have the outgoing wave
 $\left (
 \begin{array}{c}
 S_{12} \\
 S_{22}
 \end{array}
 \right ) p^{-inu+ n \ell}$.
 From (3.21) we find that the conditions for the vanishing of the
 $g(f)$\--component in the incoming wave look like
 $$
 \frac{N_-F_1^-}{N_+F_1^+} = \pm (1-p^{2 \ell +1})
 \frac{1-p^{-iu + \ell}}{1+p^{-iu+ \ell}}
 \eqno(3.23)
 $$
 (``$+$" for $g$ and ``-" for $f$).
 Using the definition above we read off the $S$\--matrix from (3.21):
 $$
 S_{KZ}(u)= \frac{c(u)}{c(-u)}
 \left (
 \begin{array}{c c}
 \frac{p^{2iu} (1-p^{2 \ell})}{1-p^{2iu+2\ell}} &
 \frac{p^\ell (1-p^{2iu})}{1-p^{2iu+2 \ell}} \\
 & \\
 \frac{p^{2iu + \ell}(1-p^{2iu})}{1-p^{2iu+2 \ell}} &
 \frac{p^{2iu}(1-p^{2 \ell})}{1-p^{2iu+2 \ell}}
 \end{array}
  \right ) .
 \eqno(3.24)
 $$
 Recalling the definition of the $R$-matrix (3.4) and its zero\--weight
 part $R_0(z)$ (3.4b), we see that (3.24) can be represented in the form
 $$
 S_{KZ} (u) =
 \left (
 \begin{array}{cc}
 0 & p^{-iu} \\
 p^{iu} & 0
 \end{array}
 \right )
 R_0 (p^{iu})U(p^{iu}) =
 $$
 $$
 = \Lambda^{-1} \sigma_1 R_0 (p^{iu}) U(p^{iu})
 \eqno(3.25)
 $$
 where $\Lambda = diag \{ p^{iu}, p^{-iu} \}$ as before (see (3.10) and
 (3.13)) and
 $$
 U(p^{iu}) = p^{iu}
 \frac{c(u)}{c(-u)}
 \eqno(3.26)
 $$
 is the scalar unitarizing factor.
 Note that $[R_0(p^{iu})$, $\Lambda^{-1} \sigma_1 ] =$
 $[S_{KZ}(u)$, $\Lambda^{-1} \sigma_1 ]= 0$.
 So the local (bare) ``$S$\--matrix" $R_0(z)$ in (3.10) (connecting two
 neighboring sites) reproduces itself in the global scattering, getting
 ``dressed" by the scalar infinite product factor $U(z)$.

 So we have completely solved the scattering problem for the difference
 operator in the l.h.s. of the q-KZ eq.~(3.10).
 The scattering eigenfunctions are precisely $F_n^\pm$ and the
 $S$\--matrix is given by eqs.~(3.24) or (3.25).
 A crucial part of the argument is the transition from the first\--order
 matrix difference equation to the pair of decoupled second\--order
 recurrence relations (3.16) that can be actually solved in terms of the
 continuous $q$\--Jacobi polynomials.
 This procedure is completely parallel to that in the ``classical" $(q
 \rightarrow 1)$ limit though much more involved technically.

 It is interesting to note that in the $q$-deformed case, the choice (3.15) of
 linear combinations leading to a reasonable pair of decoupled
 second\--order equations is {\em not unique}.
 Here by a reasonable equation we mean  one which can be actually
 solved, with the asymptotics of the solutions being known explicitly.
 The other possibility is to take (see (A26))
 $$
 \tilde{F}_n^\pm = f_n \pm p^{-2iu \pm \ell} g_n
 \eqno(3.27)
 $$
 instead of (3.15) (the classical limit is the same).
 In this case, eq.~(3.14) is equivalent to the following second order
 recurrence relations:
 $$
 \frac{1-p^{2n+4\ell -2}}{1-p^{2n+2 \ell -2}}
 \tilde{F}_{n+1}^+ +p^{2 \ell}
 \frac{1-p^{2n-2}}{1-p^{2n+2 \ell -2}}
 \tilde{F}_{n-1}^+ = p^\ell (p^{iu} + p^{-iu}) \tilde{F}_n^+
 \eqno(3.28a)
 $$
 $$
 \frac{1-p^{2n+4\ell }}{1-p^{2n+2 \ell -2}}
 \tilde{F}_{n+1}^- +p^{2 \ell}
 \frac{1-p^{2n}}{1-p^{2n+2 \ell -2}}
 \tilde{F}_{n-1}^- =
 $$
 $$
 = p^\ell (p^{iu} + p^{-iu})
 \frac{1-p^{2n}}{1-p^{2n-2}} \tilde{F}_n^-
 \eqno(3.28b)
 $$
 (for details see Appendix A).
 Now the boundary conditions at the origin are different:  $\tilde{F}_0^+$
 is finite and $\tilde{F}_1^- = 0$ (the last condition is forced by (3.28b)).
 One can see that the solutions to (3.28a) and (3.28b) can be also
 expressed through continuous $q$\--Jacobi polynomials (of a different
 type):
 $$
 \tilde{F}_n^+ = \tilde{F}_1^+ R_{n-1}^{(\ell - 1/2 , \ell - 1/2)} (1/2
 (p^{iu} + p^{-iu}); p)
 \eqno(3.29a)
 $$
 $$
 \tilde{F}_n^- = \tilde{F}_2^- \cdot
 \frac{p^{-(n-1)} - p^{n-1}}{p^{-1}-p}R_{n-2}^{(\ell + 1/2 , \ell + 1/2)}
 (1/2 (p^{iu} + p^{-iu}); p)
 \eqno(3.29b)
 $$
 From the known asymptotics of the l.h.s. of (3.29) one obtains an
 $S$\--matrix which is {\em different} from (3.24).
 This is not surprising because now we work with another type of boundary
 conditions at the origin.  When expressed in terms of $f_n$ and $g_n$
 and compared to the original choice of boundary conditions, they look
 much less natural.
 For example, they imply the  spectral parameter dependent constraint
 $g_1 = p^{2iu + \ell } f_1$.
 Nevertheless, the meaning of this extra solution to the spectral problem,
 specific to the $q$\--deformed case, deserves better understanding.
 Some more comments on this point will be made in the next Section.

 The polynomials $R_n^{(\alpha, \alpha)} (x; p)$ in (3.29) are known also
 as Rogers\--Askey\--Ismail polynomials \cite{R, AI, GR} or Macdonald
polynomials
 for root system $A_1$ \cite{M}.
 The appearance of Macdonald polynomials in the context of q-KZ equations
 was already discussed by Cherednik \cite{CHE, CH}.

 \bigskip
 \leftline{\bf 4.  The spectral problem at level zero}

 In this Section we discuss the general result of Sec.~3 in the special
 case of level zero ($k=0$ in (3.9)).
 This deserves particular attention because, as we shall see, the spectral
 problem (3.14) for this simplest case yields the physical $S$\--matrix
 for the XXZ spin $-1/2$ anti\--ferromagnet suggesting at the same time a
 nice geometrical interpretation in terms of scattering in the quantum
 group $SL_q (2, {\bf R})$.

 For $k=0 ~~ p=q^2$, $\ell = 1/2$, $u = \lambda$ (see (3.11)-(3.13)) and
 it is more convenient to use the original notation $q$ and $\lambda$.
 Specializing (3.15) and (3.18) to this case we obtain
 $$
 F_n^+ = F_1^+ R_{n-1}^{(0,1)} (\frac{1}{2} (q^{2 i \lambda} + q^{-2i
\lambda});
 q^2)
 \eqno(4.1a)
 $$
 $$
 F_n^- = F_1^- \cdot
 \frac{q^{-2n} -q^{2n}}{q^{-2} -q^2} R_{n-1}^{(1,0)}(\frac{1}{2} (q^{2 i
 \lambda}
 + q^{-2i \lambda}); q^2)
 \eqno(4.1b)
 $$
 with the asymptotics
 $$
 F_n^+ |_{n \rightarrow \infty} = N_+ F_1^+ q^{n-1}
 \left (
 \frac{q^{-2i \lambda} \Gamma_{q^4} (-i \lambda)}{(1+q^{-2i \lambda+1})
 \Gamma_{q^4} (-i \lambda + 1/2)} q^{2in \lambda} +
 \right .
 $$
 $$
 \left .
 + \frac{q^{2i \lambda} \Gamma_{q^4} (i \lambda)}{(1 +q^{2 i \lambda +1})
 \Gamma_{q^4} (i \lambda + 1/2)} q^{-2in \lambda} \right )
 \eqno(4.2a)
 $$
 $$
 F_n^- |_{n \rightarrow \infty} =\frac{ N_- F_1^- q^{n+1}}{1-q^4}
 \left (
 \frac{q^{-2i \lambda} \Gamma_{q^4} (-i \lambda)}{(1-q^{-2i \lambda+1})
 \Gamma_{q^4} (-i \lambda + 1/2)} q^{2in \lambda} +
 \right .
 $$
 $$
 \left .
 + \frac{q^{2i \lambda} \Gamma_{q^4} (i \lambda)}{(1 -q^{2 i \lambda +1})
 \Gamma_{q^4} (i \lambda + 1/2)} q^{-2in \lambda} \right )
 \eqno(4.2b)
 $$
 where
 $$
 F_n^\pm = f_n \pm q^{-2i \lambda} g_n
 \eqno(4.3)
 $$
 Had we included the factor $\rho (z)$ (3.6) in the
 $R$\--matrix, its effect would have been  to force $F_n^\pm$ to vanish at
 negative $n$ because the zeros of $\rho (z)$ lie just at the points
 $q^{2n+1}$, $n < 0$.

 The $S$-matrix (3.25) is
 $$
 S_{KZ} (\lambda ) = \Lambda^{-1} \sigma_1 R_0 (q^{2i \lambda}) \cdot q^{2
 i \lambda}
 \frac{\Gamma_{q^4} (i \lambda) \Gamma_{q^4} (-i \lambda +
 1/2)}{\Gamma_{q^4} (-i \lambda) \Gamma_{q^4} (i \lambda + 1/2)}
 \eqno(4.4)
 $$
 where
 $$
 \Lambda^{-1} \sigma_1 = \left (
 \begin{array}{c c}
 0 & q^{-2 i \lambda} \\
 q^{2 i \lambda} & 0
 \end{array}
 \right )
 \eqno(4.5)
 $$
 commutes with $R_0 (q^{2 i \lambda})$.
 Note that $F_n^\pm$ (4.3) are just the eigenvectors of $R_0 (q^{2 i
 \lambda})$.
 The eigenvalues of $S_{KZ}(\lambda )$ are given by
 $$
 S_\pm (\lambda ) = q^{4 i \lambda}
 \frac{1 \pm q^{-2i \lambda+1}}{1 \pm q^{2 i \lambda +1}} \cdot
 \frac{\Gamma_{q^4} (i \lambda) \Gamma_{q^4} (-i \lambda +
 1/2)}{\Gamma_{q^4} (-i \lambda) \Gamma_{q^4} (i \lambda + 1/2 )}
 \eqno(4.6)
 $$
 This $S$-matrix actually coincides (up to the
 trivial matrix factor $\Lambda^{-1} \sigma_1$) with the kink\--antikink
 scattering matrix $S_{k-a} (\lambda )$ for the spin$- 1/2$ XXZ model in
 the antiferromagnetic regime (with  anisotropy parameter $- \log q$):
 $$
 S_{KZ} ( \lambda ) = \Lambda^{-1} \sigma_1 S_{k-a} ( \lambda) .
 \eqno(4.7)
 $$
 For the definition of $S_{k-a} ( \lambda )$ see, for example, eq.~(6.18)
 in \cite{DFJMN}.
 The eigenvalues of $S_{k-a} ( \lambda )$
 (corresponding to scattering with a given parity) are $\pm S_\pm
 (\lambda)$.
 The role of the extra matrix factor $\Lambda^{-1} \sigma_1$ is to change
 the sign of $S_-$.
 In the classical limit, this factor reduces to the constant matrix
 $\sigma_1$ and produces the minus sign noted there.

 The $q$-Jacobi polynomials $R_n^{(\alpha , \beta )}$ are known to provide
 (for some values of $\alpha , \beta $) the full set of spherical
 harmonics on the quantum group $SL_q (2)$ \cite{NM}.
 Recalling that in this case $\alpha = |m-n|$, $\beta = |m+n|$ where
 $m(n)$ is the number of the left (right) $SO(2)$\--harmonics, we see that
 the values of $\alpha , \beta$ in (4.1) correspond to spinorial harmonics
 $( 1/2 , 1/2)$ and $(1/2, -1/2)$.
 This is in line with the Dirac\--Bargmann\--Wigner analogy.

 Thus eq.~(4.7) suggests an interpretation of the scattering of physical
 excitations in the XXZ model in terms of purely geometrical scattering of
 a spinning particle on the quantum group.
 To be more precise, we need to consider the scattering on the dual object
 to the compact real form of $SL_q(2)$.
 Indeed, our coordinate variable $n$ is just the spectral index of the
 $q$\--Jacobi polynomials.
 In the case of the $S$\--wave scattering (involving only zonal spherical
 harmonics) this dual object can be in some sense identified with a {\em
 non\--compact} real form of $SL_q(2)$.
 Analytically, this  shows up in the nice self\--duality symmetry of
 Macdonald polynomials (which play the role of zonal spherical functions on
 $SL_q(2)$) with respect to the exchange of the argument and spectral
 index \cite{FZCO, FZPL}.
 However, for spinorial harmonics this duality comes into play in a more
 non\--trivial way which is not completely clear for us at the moment.
 To illustrate this, let us consider the classical $(q \rightarrow 1)$
 limit of the $q$\--Jacobi polynomials.

 If we take the limit $q \rightarrow 1$ for fixed $\varphi = 2 \lambda \log
 q$ (so that $q^{2 i \lambda} \rightarrow e^{i \varphi})$   the $q$\--Jacobi
 polynomials
 $R_n^{(\alpha , \beta )} ( \frac{1}{2}
  (q^{2 i \lambda} + q^{-2 i \lambda}); q^2)$
 go to the classical Jacobi polynomials $P_n^{\alpha , \beta )} (\cos \varphi
 )$ up to normalization.
 They yield the restriction of spherical harmonics on the compact real
 form of $SL(2)$ to the maximal torus parametrized by the coordinate $\varphi$.
 The scattering in the index $n$ {\em after} taking this limit gives a
 trivial $S$\--matrix having nothing to do with the $S_{k-a} (\lambda )$ at
 $q=1$ (2.12).
 This means that the limits $n \rightarrow \infty$ and $q \rightarrow 1$
 do not commute.
 To achieve agreement with the $S$\--matrix (2.12) we need another
 classical limit!
 This is $q \rightarrow 1$ and $n \rightarrow \infty$ for fixed $x \sim n
 \log q$ (now $x$ is a continous variable) and $\lambda$.
 Then, by considering the asymptotics {\em at large x} we recover the
 right XXX $S$\--matrix (2.12).
 In this case the $q$\--Jacobi polynomials go to Legendre (or Gegenbauer)
 functions $P_{- \alpha - 1/2 + i \lambda} (\cosh x)$ and {\em not} to
 Jacobi functions (as one would expect).
 This can be easily seen from (B6) and was already pointed out by
 Koornwinder \cite{K}.

 Finally, in the case $k=0$
 the spectral problem for $\tilde{F}_n^+$ (3.28a) gives yet another
 eigenvalue of the full XXZ $S$\--matrix corresponding to {\em
 kink\--kink} (or antikink\--antikink) scattering.
 this is just the result obtained in our earlier papers \cite{FZCO, FZPL} by
 considering the scalar spectral problem on a quantum hyperbolic plane.
 This eigenvalue turns out to be equal to $U(q^{2 i \lambda})$ (3.26).

 \bigskip
 \leftline{\bf 5.  The limit of infinite level}

 The limit of infinite level $k \to \infty$ is obscure in the original
 version of q-KZ eqs.~(3.9) because $q^{k+2}=p$ goes to zero.
 However, the form (3.14) is more appropriate for taking this limit since
 $p$ does not appear in the shift operator explicitly and $p^\ell =
 p^{1/(k+2)} =q$ is fixed.

 Taking the limit $k \to \infty$ in (3.14) we get
 $$
 \left (
 \begin{array}{c}
 f_{n+1} \\ g_{n+1}
 \end{array}
 \right ) =
 \left (
 \begin{array}{c c}
 q^{-2i \lambda} & 0 \\
 0 & q^{2i \lambda}
 \end{array}
 \right)
 \left (
 \begin{array}{c c}
 q & 1-q^2 \\
 0 & q
 \end{array}
 \right )
 \left (
 \begin{array}{c}
 f_n \\ g_n
 \end{array}
 \right )
 , n \geq 1
 \eqno(5.1)
 $$
 The boundary conditions are imposed by fixing the values of $f_1 , g_1$.
 The values of $f_0 , g_0$ are inessential (see the discussion of this
 point in Sec.~3).
 The second matrix on the r.h.s. of (5.1) is just the
 zero\--weight piece of the inverse of the familiar
constant $R$\--matrix for $U_q(sl_2)$ in the
 fundamental representation.
 The appearance of $U_q(sl_2)$ in this context looks quite natural since
 the limit of infinite level usually means ``forgetting the affinization"
 of affine (quantum) algebras and thus leads to finite\--dimensional Lie
 algebras.

 On the other hand, we will presently show that for the particular values
 of $q^2 = P^{-1}$ where $P$ is a prime number (5.1) has a nice
 interpretation in terms of harmonic analysis on the $P$\--adic group
 $SL(2, {\bf Q}_P)$.
 [In view of the extensive use above of the symbol $p$ as defined in
 eq.~(3.11), we are straying here from number theoretic custom and denote
 a prime by $P$].
 To see this, let us derive from (5.1) the second order recurrence
 relations for
 $F_n^\pm = f_n \pm q^{-2i \lambda} g_n$ and  for $\tilde{F}_n^\pm = f_n \pm
 q^{-4 i \lambda \pm 1}g_n$.
 From (3.16) and (3.28) we obtain:
 $$
 P F_{n+1}^\pm + F_{n-1}^\pm = 2 x P^{1/2} F_n^\pm + \delta_{n,1}
 (F_0^\pm \mp F_1^\pm ); n \geq 1
 \eqno(5.2)
 $$
 $$
 \left \{
 \begin{array}{l l r}
 P \tilde{F}_{n+1}^+ + \tilde{F}_{n-1}^+ = 2 x P^{1/2} \tilde{F}_n^{+}+
 \delta_{n,1} (\tilde{F}_0^+ - \tilde{F}_2^+ ) ; &
 n \geq 1 & ~~~~(5.3a) \\
 && \\
 P \tilde{F}_{n+1}^- + \tilde{F}_{n-1}^- = 2 x P^{1/2} \tilde{F}_n^-; &
 n \geq 2 , \tilde{F}_1^- = 0 . & ~~~~(5.3b)
 \end{array}
 \right .
 $$

 Consider the equation (5.3a) first.
 One easily recognizes it as the recurrence relation for the
 Mautner\--Cartier polynomials \cite{MC} which are zonal spherical
 functions on the $P$\--adic symmetric space $H_P = PGL (2, {\bf
 Q}_P)/PGL$$ (2, {\bf Z}_P)$ of the group $PGL (2, {\bf Q}_P ) $ $= GL (2,
 {\bf Q}_P)/ {\bf Q}_P^*$ where $PGL(2, {\bf Z}_P)$ is its maximal compact
 subgroup.
 The $\delta$\--symbol term in (5.3a) provides the proper boundary
 condition.
 The space $H_P$ is known as a Bruhat\--Tits tree \cite{BT} and it can be
 represented as the homogeneous tree with each vertex being joined to
 $P+1$ ``neighboring" vertices by edges (Fig.~1).
 The $PGL(2, {\bf Q}_P)$\--invariant Laplacian acting on scalar\--valued
 functions of the vertices can be naturally defined as a sum of the values
 in all nearest neighbors of a given vertex minus the value of the
 function at this vertex times the number of the nearest neighbors ($P+1$
 in this case).
 One can arbitrarily choose a ``central" vertex of the tree and consider
 zonal spherical functions defined by the two conditions:  1) they are
 eigenfunctions of the Laplacian; 2) they are ``spherically symmetric",
 i.e., take one and the same value at all points equally spaced from the
 center.
 Calling $\varphi_n$ the value of the function of the vertices on distance
 $n+1$ from the center, one can easily write down the eigenvalue equation
 (Fig.~1):
 $$
 P \varphi_{n+1} + \varphi_{n-1} = (E + P + 1) \varphi_n , ~ n \geq 2
 $$
 $$
 (P+1) \varphi_2 = (E + P + 1) \varphi_1,
 \eqno(5.4)
 $$
 where $E$ is the eigenvalue.
 Introducing the new spectral variable $x = 1/2 (P^{i \lambda} + P^{-i
 \lambda})$ by
 $$
 E + P + 1 = 2 x P^{1/2}
 \eqno(5.5)
 $$
 we see that (5.3a) and (5.4) coincide.

 The Mautner-Cartier polynomials are $P$\--adic analogs of the Legendre
 (or Gegenbauer) functions.
 The polynomials $F_n^\pm(x)$ in (5.2) should thus provide a $P$\--adic
 analog of the Jacobi\--functions with particular values of $\alpha$ and
 $\beta$.
 Actually, (5.3a) and (5.2) differ only by the boundary condition at the
 origin.

 We now suggest an ``arboreal''
 interpretation of the equation (5.2) for $F_n^+$
 similar to that for $\tilde{F}_n^+$ described above.
 Suppose we attach the values $\chi_n$ to the edges of the tree rather
 than to the vertices so that this function is again spherically
 symmetric:  its values depend only on the distance $n$ of the edges to
 the center $C$ (Fig.~1).
 The definition of the Laplacian remains the same as above with the change
 of vertices to edges, the nearest neighbors of an edge being all the
 edges having a common end with it (there are $2P$ of them).
 Now the eigenvalue equation looks like
 $$
 P \chi_{n+1} + \chi_{n-1} + (P-1) \chi_n = (E + 2P) \chi_n ~, ~~~n \geq 2
 $$
 $$
 P \chi_2 + P \chi_1 = (E+2P) \chi_1
 \eqno(5.6)
 $$
 Recalling (5.5) this can be rewritten as
 $$
 P \chi_{n+1} + \chi_{n-1} = 2 x P^{1/2} \chi_n ~ , ~~~ n \geq 2
 $$
 $$
 P \chi_2 + \chi_1 = 2 x P^{1/2} \chi_1
 \eqno(5.7)
 $$
 that is exactly the equation (5.2) for $F_n^+$.

 Keeping in mind the analogy with the Dirac equation, we can say that the
 $R$\--matrix for $U_q (sl_2)$ in (5.1) provides a ``square root" of the
 Laplace operator on the $P$\--adic tree $(q^2 = P^{-1})$.
 This is one more face of the $P$\--adics\--quantum group connection
 \cite{M, F, Z, FZCO}.
 This opens a way to introduce spinorial harmonics on $P$\--adic groups
 that would be very interesting from various points of view.

 \bigskip
 \leftline{\bf 6.  Discussion}

 In considering the spectral problem for the q-KZ equation, we were
 motivated by the scalar spectral problem for the Laplace operator on a
 curved (quantum) space.
 The latter was known to yield the scattering phase for the kink\--kink
 scattering in the spin$-1/2$ XXZ antiferromagnet \cite{Z, FZCO, FZPL}
 Since the physical excitations in this model are spin $-1/2$ kinks it is
 natural to expect that their scattering matrix would come from the
 spectral problem for a matrix first\--order operator on the same quantum
 space whose ``square" gives the Laplace operator in the spirit of Dirac's
 trick.
 The difference operator appearing in q-KZ equations is a natural
 candidate for this.
 We have shown that this is indeed the case and the physical $S$\--matrix
 of the XXZ\--model can be  derived this way (for the q-KZ equation at zero
 level).
 However, this correspondence is perfect only for the 2\--dimensional
 zero\--weight (zero z\--projection of spin) subspace of the linear space
 ${\bf C}^2 \otimes {\bf C}^2$ on which  the solutions of the q-KZ equation
 take values.
 In terms of the XXZ\--model, this corresponds to the kink\--antikink
 scattering.
 As for the kink\--kink or antikink\--antikink channels, they seem to have
 nothing to do with the $(++)$ and $(--)$ components of the q-KZ
 solutions, since the latter have  trivial scattering.
 We found instead that the phase shifts in these channels can be obtained
 from the scattering problem for the same q\--KZ operator with another
 type of boundary conditions at the origin.
 The conceptual understanding of this situation is obscure.

 Anyway, we succeeded in describing at least the kink\--antikink
 scattering in the spin$-1/2$ XXZ antiferromagnet in terms of particular
 continuous q\--Jacobi polynomials which yield the spinorial harmonics on
 the quantum group $SL_q(2)$.
 This picture is in good agreement with the conjecture made in our earlier
 paper \cite{FZK} that the scattering processes in integrable systems
 are of a {\em purely geometric} nature.

 The above connection with the XXZ model comes already at the level $k$
 equal to zero.
 In Sec.~3 we solved the scattering problem for the
$U_q(\widehat{sl_2})$\--q-KZ
 equation in the fundamental representation for arbitrary level.
 Does this have any interpretation in terms of excitation scattering in
 integrable models?
 Based on the results for the corresponding scalar spectral problem
 \cite{FZCO} we can conjecture that this case is related to generalized
 $SL_q(k+2)$\--magnetics, or
 equivalently, to the ${\bf Z}_{k+2}$ Baxter statistical model on
 the square lattice.

 A generalization to q-KZ equations for multipoint functions would also be
 of interest.
 It should correspond to the multiparticle scattering in integrable models
 which is known to factorize.

Matsuo \cite {MA2} has found a Jackson-integral representation of the
Jordan-Pochhammer type for solutions of the $q-$KZ equation.
Combining this with our results, one could obtain a new Jackson integral
representation for the $q$-Jacobi polynomials.

 Finally, let us remark on the
 relationship between the monodromy (or connection) problem on the one
 hand and  the
 scattering problem on the other hand.
 Our results suggest that they are in some sense dual to each other.
 The usual setting of the monodromy (connection) problems for q-KZ
 equations is quite the opposite to what we have done:  the spectral
 parameter $\lambda$ is fixed once and for all, and one compares the
 solutions regular at $z=0$ with those regular at $z = \infty$ (in the
 variable $n$ this corresponds to $n = - \infty$ and $n = \infty$).
 Then the physical $S$\--matrices appear as the ratios of two special
 solutions and they are now functions of $z$, not $\lambda$.
 This indicates a remarkable duality between the coordinate variable and
 the spectral parameter.

 \bigskip
 \leftline{\bf Acknowledgments}

We thank A. Gorsky, S. Shatashvili and P. Wiegmann for discussions
and to I. Cherednik and T. Koornwinder for sending us their papers. One of
us (A.Z.) wishes to thank Prof. J. Peter May for the hospitality of the
Mathematical Disciplines Center at the University of Chicago.

\newpage
\centerline{\bf Appendix A}

In this Appendix we show how one can find the specific linear
combinations of $f_n , g_n$ satisfying the equation for $q$\--Jacobi
polynomials.

Let us write the q-KZ equation (3.14) in the form
$$
\left (
\begin{array}{c}
f_{n+1} \\
g_{n+1}
\end{array}
\right ) =
\left (
\begin{array}{c c}
a_n & b_n \\
c_n & d_n
\end{array}
\right )
\left (
\begin{array}{c}
f_n \\ g_n
\end{array}
\right )
\eqno(A1)
$$
It is clear that $f_n$ and $g_n$ satisfy decoupled second\--order
recurrence relations but generally these are not very helpful.
Our task is to extract an equation which can be solved!
In the limit $q=1$ the situation is quite similar though much simpler:
$f(x)$ and $g(x)$ themselves obey two decoupled second order differential
equations of rather complicated form but, fortunately, the ``right" linear
combinations $f \pm g$ leading to the familiar hypergeometric equations
are more or less obvious from the very beginning.

The idea is to apply to (A1) a similarity transformation in such a way
that the resulting second order equations would be as simple as possible.
So let us take a non\--degenerate constant matrix $U = \{ u_{ij} \}$;
$i,j = 1,2$ and apply it to (A1).
We get
$$
\left (
\begin{array}{c}
\tilde{f}_{n+1}   \\
\tilde{g}_{n+1}
\end{array}
\right )
= \left (
\begin{array}{c c}
\tilde{a}_n & \tilde{b}_n \\
\tilde{c}_n & \tilde{d}_n
\end{array}
\right )
\left (
\begin{array}{c}
\tilde{f}_n \\ \tilde{g}_n
\end{array}
\right )
\eqno(A2)
$$
where
$$
\left (
\begin{array}{c}
\tilde{f}_n \\ \tilde{g}_n
\end{array}
\right )
= U^{-1}
\left (
\begin{array}{c}
f_n \\ g_n
\end{array}
\right ) ~~,
\eqno(A3)
$$
$$
\left (
\begin{array}{c c}
\tilde{a}_n & \tilde{b}_n \\
\tilde{c}_n & \tilde{d}_n
\end{array}
\right )
= U^{-1}
\left (
\begin{array}{c c}
a_n & b_n \\
c_n & d_n
\end{array}
\right )
U
\eqno(A4)
$$
The matrix equation (A2) is equivalent to the pair of decoupled
second order equations for $\tilde{f}_n$ and $\tilde{g}_n$:
$$
\tilde{b}_{n-1} \tilde{f}_{n+1} + \tilde{b}_n \Delta_{n-1}
\tilde{f}_{n-1} = (\tilde{a}_n \tilde{b}_{n-1} + \tilde{d}_{n-1}
\tilde{b}_n ) \tilde{f}_n
\eqno(A5a)
$$
$$
\tilde{c}_{n-1} \tilde{g}_{n+1} + \tilde{c}_n \Delta_{n-1}
\tilde{g}_{n-1} =
(\tilde{d}_n \tilde{c}_{n-1} + \tilde{a}_{n-1} \tilde{c}_n ) \tilde{g}_n
\eqno(A5b)
$$
where
$$
\Delta_n = \det R_0 ( p^{n+ \ell}) =
\frac{p^{2 \ell} (1-p^{2n})}{1-p^{2n+4 \ell}}
\eqno(A6)
$$
The matrix elements in (A4) have the following general form
$$
\tilde{a}_n = \frac{\alpha_0 - \alpha_1 p^{2n}}{1-p^{2n+4 \ell}} ~~ ,~~~~~
\tilde{b}_n = \frac{\beta_0 - \beta_1 p^{2n}}{1-p^{2n+4 \ell}}
$$
$$
\tilde{c}_n = \frac{\gamma_0 - \gamma_1 p^{2n}}{1-p^{2n+4 \ell}} ~~ ,~~~~~
\tilde{d}_n = \frac{\delta_0 - \delta_1 p^{2n}}{1-p^{2n+4 \ell}}
\eqno(A7)
$$
where the full $n$-dependence  is indicated explicitly.
Calling $y = p^{iu}$ for brevity, we have from (A4):
$$
\alpha_0 = (\det U)^{-1} [y^{-1} p^\ell u_{11} u_{22} + y^{-1} (1-p^{2
\ell}) u_{21} u_{22} - yp^\ell u_{12} u_{21} ]
$$
$$
\alpha_1 = (\det U)^{-1} [y^{-1} p^{3 \ell} u_{11} u_{22} + y (1-p^{2
\ell}) p^{2 \ell} u_{11} u_{12} - yp^{3 \ell} u_{12} u_{21}]
$$
$$
\beta_0 = (\det U)^{-1} [y^{-1} p^\ell u_{12} u_{22} + y^{-1} (1-p^{2
\ell}) u_{22}^2 -yp^\ell  u_{12} u_{22} ]
$$
$$
\beta_1 = (\det U)^{-1} [y^{-1} p^{3 \ell} u_{12} u_{22} + y (1-p^{2
\ell}) p^{2 \ell} u_{12}^2  - yp^{3 \ell} u_{12} u_{22}]
$$
$$
\gamma_0 = (\det U)^{-1} [-y^{-1} p^\ell u_{11} u_{21} - y^{-1} (1-p^{2
\ell}) u_{21}^2 + yp^\ell u_{11} u_{21} ]
$$
$$
\gamma_1 = (\det U)^{-1} [-y^{-1} p^{3 \ell} u_{11} u_{21} - y (1-p^{2
\ell}) p^{2 \ell} u_{11}^2 + yp^{3 \ell} u_{11} u_{21}]
$$
$$
\delta_0 = (\det U)^{-1} [-y^{-1} p^\ell u_{12} u_{21} - y^{-1} (1-p^{2
\ell}) u_{21} u_{22} + yp^\ell u_{11} u_{22} ]
$$
$$
\delta_1 = (\det U)^{-1} [-y^{-1} p^{3 \ell} u_{12} u_{21} - y (1-p^{2
\ell}) p^{2 \ell} u_{11} u_{12} + yp^{3 \ell} u_{11} u_{22}]
\eqno(A8)
$$
The equations (A5) are now written as:
$$
(1-p^{2n+4 \ell})( \beta_0 - \beta_1 p^{2n-2}) \tilde{f}_{n+1} + p^{2
\ell} (1-p^{2n-2}) (\beta_0 - \beta_1 p^{2n}) \tilde{f}_{n-1} =
$$
$$
= \left ( ( \alpha_0 - \alpha_1 p^{2n}) (\beta_0 - \beta_1 p^{2n-2}) +
(\delta_0 - \delta_1 p^{2n-2}) ( \beta_0 - \beta_1 p^{2n}) \right )
\tilde{f}_n
\eqno(A9a)
$$
and
$$
(1-p^{2n+4 \ell})( \gamma_0 - \gamma_1 p^{2n-2}) \tilde{g}_{n+1} + p^{2
\ell} (1-p^{2n-2}) (\gamma_0 - \gamma_1 p^{2n}) \tilde{g}_{n-1} =
$$
$$
= \left ( ( \delta_0 - \delta_1 p^{2n}) (\gamma_0 - \gamma_1 p^{2n-2}) +
(\alpha_0 - \alpha_1 p^{2n-2}) ( \gamma_0 - \gamma_1 p^{2n}) \right )
\tilde{g}_n
\eqno(A9b)
$$

We are going to compare (A9) with the recurrence relation (B8) for
$q$\--Jacobi polynomials (see Appendix B below).
For general values of $(\alpha , \beta )$ they look quite different.
However, for particular values: I) $\beta = \alpha + 1$;
II) $\beta = \alpha - 1$;
III) $\beta = \alpha$, considerable simplifications in (B8) occur.
For example, in the case I) we have
$$
(1-p^{2n+4 \alpha +4})
(1-p^{2n+ 2 \alpha +1})
R_{n+1} (x) + p^{2 \alpha +1} (1-p^{2n}) (1-p^{2n+2 \alpha +3}) R_{n-1}
(x) =
$$
$$
= - (1-p^2) (1-p^{2 \alpha +1})p^{2n+2 \alpha +1} R_n (x) + 2 x p^{\alpha
+ 1/2} (1-p^{2n + 2 \alpha +3} )
$$
$$
(1-p^{2n+ 2 \alpha +1}) R_n (x) ,
\eqno(A10)
$$
whose l.h.s. is {\em identical} to that of (A9a) provided
$$
\alpha = \ell - 1/2
\eqno(A11)
$$
$$
\beta_1 = \beta_0 p^{2 \ell}
\eqno(A12)
$$
$$
\tilde{f}_n = R_{n-1} (x).
\eqno(A13)
$$
What about the right-hand sides?
A straightforward calculation shows that they are identical provided
$$
\left \{
\begin{array}{c r}
\alpha_0 + \delta_0 = 2 x p^{\alpha + 1/2} & ~~~~~(A14) \\
& \\
\alpha_1 + \delta_1 = 2xp^{3 \alpha + 3/2} & ~~~~~(A15) \\
& \\
\alpha_0 - p^{-2 \ell} \alpha_1 = 1 - p^{2 \ell} . & ~~~~~(A16)
\end{array}
\right .
$$
One can see that all these conditions are mutually consistent.

They determine the matrix elements of $U$.
Solving (A11)-(A16) using (A8), we obtain the following simple relations:
$$
2x = y+y^{-1} = p^{i u} + p^{-iu}
\eqno(A17)
$$
$$
\frac{u_{12}}{u_{22}} = - y^{-1} = - p^{-iu}
\eqno(A18)
$$
The other matrix elements of $U$ are not fixed yet because so far we have
used only one equation from the pair (A9).

It turns out that the equation for $\tilde{g}_n$  (A9b) can be obtained
from (B8) in a similar way.
Namely, let us rewrite (B8) in the case II $(\beta ' = \alpha  ' - 1$;
the primes indicate that the values of $\alpha$ and $\beta$ may be
different from the case above)
as equations for the polynomials $\tilde{R}_n (x)$
normalized as follows:
$$
\tilde{R}_n (x) = p^{-n} (1-p^{2n+2 \alpha ' - 2})
R_{n-1}^{(\alpha ' , \alpha ' - 1)} (x; p)~~.
\eqno(A19)
$$
We arrive at
$$
(1-p^{2n+4 \alpha ' -2})
(1-p^{2n+ 2 \alpha ' - 3})
\tilde{R}_{n+1} (x)
$$
$$
+ p^{2 \alpha  ' -1}(1-p^{2n - 2}) (1-p^{2n+2 \alpha ' -1}) \tilde{R}_{n-1}
(x) =
$$
$$
=  (1-p^2) (1-p^{2 \alpha  ' -1})p^{2n+2 \alpha ' -3} \tilde{R}_{n} (x)+
$$
$$
2 x p^{\alpha ' - 1/2} (1-p^{2n + 2 \alpha '-1} )
(1-p^{2n+ 2 \alpha ' -3}) \tilde{R}_n (x) ~~.
\eqno(A20)
$$
The expressions on the l.h.s.of (A9b) and (A20) coincide, provided
$$
\alpha ' = \ell + 1/2
\eqno(A21)
$$
$$
\gamma_1 = \gamma_0 p^{2 \ell}
\eqno(A22)
$$
$$
\tilde{g}_n = \tilde{R}_n(x)
\eqno(A23)
$$
Note that the values of $\alpha_i , \delta_i$ are already fixed by
(A14)-(A16).
Substituting them into the r.h.s. of (A9b) and taking into account
(A21)-(A23) one obtains exactly the r.h.s. of (A20).
(A22) gives an extra relation for the matrix elements of $U$:
$$
\frac{u_{21}}{u_{11}} = y = p^{iu}
\eqno(A24)
$$
Now, recalling (A3), we find from (A18) and (A24) the desired linear
combinations:
$$
F_n^\pm = f_n \pm p^{-iu} g_n
\eqno(A25)
$$

Now to the case III $(\beta = \alpha )$.
The recurrence relation for $R_n^{(\alpha , \alpha )}$ has the form (B12)
that is even simpler than (A10) or (A20).
One can try to find another similarity transformation to fit the q-KZ
equations (A1) to a pair of the equations like (B12) with different
values of $\alpha$.
Remarkably, it turns out to be possible, with the new linear combinations
being
$$
\tilde{F}_n^\pm = f_n \pm p^{\pm \ell - 2 iu} g_n ~~.
\eqno(A26)
$$
The calculations are quite similar to those above.

\newpage
\centerline{\bf Appendix B}

Here we collect some necessary formulas related to $q$\--Jacobi
polynomials and their asymptotic properties.
As in the main text, we denote the base parameter as $p$ rather than $q$
to distinguish it form the deformation parameter of the quantum group.

It is convenient to introduce $q$\--Jacobi polynomials as a particular
case of a very general family of Askey\--Wilson polynomials explicitly
given by
$$
p_n (\frac{1}{2} (z+z^{-1}); a, b, c, d |p) =
(ab; p)_n (ac; p)_n (ad, p)_n a^{-n} ~~.
$$
$$
\cdot_4 \phi_3 \left [
\begin{array}{c r}
p^{-n}, abcdp^{n-1} ,~ az,~ az^{-1} & \\
& ; p, p \\
ab,~ ac,~ ad & \\
\end{array}
\right ]
\eqno(B1)
$$
where
$$
_4 \phi_3 \left [
\begin{array}{l r}
\alpha ,~ \beta ,~ \gamma , ~\delta & \\
& ; p , ~ x \\
\lambda , ~ \mu , ~ \nu & \\
\end{array}
\right ]
= \sum_{k = 0}^\infty
\frac{(\alpha ; p)_k (\beta ; p)_k ( \gamma ; p)_k (\delta ; p)_k
}{(\lambda ;p)_k (\mu ; p)_k (\nu ; p)_k (p ; p)_k} x^k
\eqno(B2)
$$
is the basic hypergeometric function.
The $p_n$ are polynomials in $ (z+z^{-1})/2$ of degree $n$.
An important property, not obvious from the definition, is that they are
actually symmetric with respect to permutations of the parameters $a, ~b,
{}~c,~d$.
We suppose that all these parameters have moduli less than $1$;
in this case the Askey\--Wilson polynomials are known to be orthogonal
for $z$ on the unit circle with a continuous measure.
We need the asymptotics of the Askey\--Wilson polynomials for large $n$.
The formula is relatively simple though the derivation is extremely
cumbersome \cite{GR, R1}:
$$
p_n (\frac{1}{2} (z+z^{-1}); a, ~b, ~c, ~d |p)
\stackrel{\sim}{\scriptstyle{n \to \infty}} z^n B (z^{-1})
+ z^{-n} B(z)
$$
$$
+ {\rm exponentially ~~ small ~~ terms}
\eqno(B3)
$$
where
$$
B(z) =
\frac{(az; p)_\infty (bz; p)_\infty (cz; p)_\infty (dz; p)_\infty}{(z^2;
p)_\infty}
\eqno(B4)
$$
Another commonly used normalization of $p_n$'s is obtained by
disregarding $z$\--independent factors in front of $_4\phi_3$ in (B1).
Though this normalization is also convenient, the symmetry in $a, ~ b,
{}~ c, ~d$ is lost in this case.
We define continuous $q-$Jacobi polynomials (they are called ``continuous"
\cite{AW} because of their orthogonality with respect to a continuous
measure, as opposed
to the so-called "little" or "big" $q$\--Jacobi polynomials which are
different $q$\--analogs of Jacobi polynomials, orthogonal with
respect to discrete measures),
using this asymmetric normalization and choosing $a, b, c, d$
as follows.

$$
a = p^{\alpha + 1/2},~~ b = - p^{\beta + 1/2},~~
c = p^{1/2} , ~~ d = - p^{1/2}
\eqno(B5)
$$
The parameters $\alpha$ and $\beta$ are supposed to be real and greater
than $-1/2$.
In terms of the basic hypergometric series, we have from (B1):
$$
R_n^{(\alpha , \beta) } (\frac{1}{2} (z+z^{-1}); p) = ~_4\phi_3
\left [
\begin{array}{l r}
p^{-n}, ~p^{n+ \alpha + \beta + 1} , ~p^{1/2} z , ~ p^{1/2} z^{-1} & \\
& ; p, ~p \\
p^{\alpha + 1} , ~ -p^{\beta +1} , ~ -p & \\
\end{array}
\right ]
\eqno(B6)
$$
Our normalization (B6) is different from  Rahman's
\cite{R2}  commonly used normalization.
For the reader's convenience we give the explicit formula connecting our
$q$\--Jacobi polynomials $R_n^{(\alpha , \beta )}$ with $P_n^{(\alpha ,
\beta )}$ defined by Rahman:
$$
R_n^{(\alpha , \beta )} (\frac{1}{2} (z+z^{-1}); p) =
\frac{(p; p)_n (-p; p)_n^2 p^{n \alpha}}{(p^{2 \alpha + 2}, p^2 )_n
(-p^{\alpha + \beta +1}; p)_n} \cdot
P_n^{(\alpha , \beta )} (\frac{1}{2} (z+z^{-1}); p)
\eqno(B7)
$$
These polynomials satisfy a three\--term recurrence relation which can be
written in the form:
$$
(1-p^{2n+2 \alpha + 2 \beta + 2}) (1-p^{2n+2 \alpha +2})(1-p^{2n+ \alpha
+ \beta})R_{n+1}^{(\alpha , \beta )} (x; p) +
$$
$$
+ p^{2 \alpha + 1} (1-p^{2n}) (1-p^{2n + 2 \beta}) (1-p^{2n + \alpha
+ \beta + 2})
R_{n-1}^{(\alpha , \beta )} (x; p) =
$$
$$
= (1+p) (p^\beta (1+p^{2\alpha}) -p^\alpha
(1+p^{2 \beta}))
p^{2n+ \alpha +1}(1-p^{2n+ \alpha + \beta +1})
R_n^{(\alpha , \beta )}(x;p) +
$$
$$
+ 2 x p^{\alpha + 1/2} (1-p^{2n+ \alpha + \beta})(1-p^{2n+ \alpha +
\beta + 1})(1-p^{2n+ \alpha + \beta +2})R_n^{(\alpha , \beta )} (x; p)
\eqno(B8)
$$
Their asymptotics at large $n$ is given by (see (B3)):
$$
R_n^{(\alpha , \beta )} (\frac{1}{2} (z+z^{-1}); p)
\stackrel{\sim}{\scriptstyle{n \to \infty}} N_{\alpha \beta}
p^{n (\alpha + 1/2 )} (z^n C_{\alpha \beta
}(z^{-1}) + z^{-n} C_{\alpha \beta}(z))
\eqno(B9)
$$
where
$$
C_{\alpha \beta} (z) =
\frac{(zp^{\alpha + 1/2}; p)_\infty (-zp^{\beta + 1/2};p)_\infty}{(z^2;
p^2)_\infty} ~~,
\eqno(B10)
$$
$$
N_{\alpha \beta}^{-1} = (p^{2 \alpha + 2}; p^2)_\infty (-p^{\alpha +
\beta +1}; p)_\infty
\eqno(B11)
$$

In the special case $\beta = \alpha$ the polynomials (B6) become
Macdonald polynomials \cite{M} for the root system $A_1$.
In the theory of $q$\--hypergeometric functions they are known as
Rogers\--Askey\--Ismail polynomials \cite{AI, GR}.
The recurrence relation (B8) then simplifies to
$$
\frac{1-p^{2n + 4 \alpha +2}}{1-p^{2n + 2 \alpha +1}} R_{n+1}^{(\alpha ,
\alpha )} (x) + p^{2 \alpha +1}
\frac{1-p^{2n}}{1-p^{2n+2 \alpha +1}} R_{n-1}^{(\alpha , \alpha )} (x) =
2xp^{\alpha + 1/2} R_n^{(\alpha , \alpha )} (x)
\eqno(B12)
$$
The famous Macdonald's parameters $q_M , ~ t_M$ are then
$$
q_M = p^2
\eqno(B13)
$$
$$
t_M = p^{2 \alpha +1}
\eqno(B14)
$$
In our normalization we have
$$
R_0^{(\alpha , \alpha )} (x) = R_n^{(\alpha , \alpha )} (\frac{1}{2}
(p^{\alpha +
1/2} + p^{- \alpha - 1/2})) = 1 ~~.
\eqno(B15)
$$
Note that unlike in the case of usual boundary conditions for three term
recurrence relations $(R_{-1} = 0 , ~R_0 = 1)$, we do {\it not} have to fix
$R_{-1}$, because the coefficient in front of it in (B12) is zero.
The same is true for the general $q$\--Jacobi polynomials (B6).

\newpage
\bigskip
\begin{center} {\bf Figure Caption}
\end{center}

{\bf Fig. 1.}  Bruhat-Tits tree $(P=2)$ with central vertex $C$.

\end{document}